\documentclass[prl,aps,twocolumn]{revtex4}

\usepackage{epsf}

\begin{document}

\title{%
\hfill{\normalsize\vbox{%
\hbox{\rm March 2002}
\hbox{\rm JLAB-THY-02-07}
\hbox{\rm SUNY-NTG-02-4}
\hbox{\rm SU-4252-755} }}\\
\vspace{-0.5cm}
{\bf Vector Meson Dominance Model for \\
Radiative Decays Involving Light Scalar Mesons}}
\author{\addtocounter{footnote}{1}
{\bf Deirdre Black}$^{\rm(a),\fnsymbol{footnote}}$}
\author{\addtocounter{footnote}{1}
{\bf Masayasu Harada}$^{\rm(b),\fnsymbol{footnote}}$}
\author{\addtocounter{footnote}{1}
{\bf Joseph Schechter}$^{\rm(c),\fnsymbol{footnote}}$}
\affiliation{$^{\rm(a)}$ Theory Group, Jefferson Lab, 12000 Jefferson
Ave., Newport News, VA 23606.}
\affiliation{$^{\rm(b)}$ Department of Physics and Astronomy,
SUNY at Stony Brook, Stony Brook, NY 11794}
\affiliation{$^{\rm(c)}$ Department of Physics, Syracuse University
Syracuse, NY 13244-1130}

\begin{abstract}
We study a vector dominance model which predicts a fairly large number
of currently interesting decay amplitudes of the types $S \rightarrow
\gamma \gamma$, $V \rightarrow S \gamma$ and $ S \rightarrow
V \gamma$,
 where $S$ and $V$ denote scalar and vector mesons,
in terms of three parameters. As an
application, the model makes it easy to study in detail a recent proposal
to boost the ratio  
$\Gamma (\phi \rightarrow f_0
\gamma)/ \Gamma(\phi \rightarrow a^0_0 \gamma)$ 
by including the isospin violating $a^0_0$-$f_0$ mixing.
However we find that this effect is actually small in our model.
\end{abstract}

\maketitle

There is increasing interest in a possible  
 nonet of light scalar mesons
(all of mass $< 1\,\mbox{GeV}$).
In addition to the well established $f_0(980)$ and
$a_0(980)$ evidence of both experimental and theoretical 
nature for a very broad 
$\sigma$ ($\simeq560$)
and  a very broad $\kappa$ 
($\simeq900$)
has been presented~\cite{kyotoconf}.
The latter two resonances are difficult to identify cleanly because
they appear to be of non Breit-Wigner type, signaling strong
interference with the non-resonant background.

Such a nonet would most likely represent meson states more complicated
than quark-anti quark type and hence would be of great importance for a
full understanding of QCD in its non-perturbative low energy regime.

Clearly it is important to study the properties of the $f_0(980)$ and
$a_0(980)$ from the point of view of how they fit into a putative
nonet family.
In particular, the reactions $\phi \rightarrow f_0 \gamma$ and 
$\phi \rightarrow a_0 \gamma$ have recently been observed
\cite{recentexpts} with good
accuracy and are considered as useful probes of scalar properties.
The theoretical analysis was initiated by 
Achasov and Ivanchenko~\cite{Achasov-Ivanchenko} and followed up by
many others~\cite{radiative}.
The models employed are essentially variants of the single $K$ meson
loop diagram to which a $\phi$-type vector meson, a photon
 and two pseudoscalars or a scalar
are attached.

In the present note we introduce a complementary approach which
emphasizes the ``family'' or symmetry aspects of the analysis.
This enables us to study the correlations
among a fairly large number of related radiative amplitudes in terms
of a few parameters, without making a commitment to a particular quark
structure for the scalars.

Our framework is that of a standard non-linear chiral Lagrangian
containing, in addition to the pseudoscalar nonet matrix field $\phi$,
the vector meson nonet matrix $\rho_\mu$ and a scalar nonet matrix
field denoted $N$.
Under chiral unitary transformations of the three light quarks;
$q_{\rm L,R} \rightarrow U_{\rm L,R} \cdot q_{\rm L,R}$, the chiral
matrix $U = \exp ( 2 i \phi/F_\pi)$,
where $F_\pi \simeq 0.131\,\mbox{GeV}$, transforms as 
$U \rightarrow U_{\rm L}\cdot U \cdot U_{\rm R}^\dagger$.
The convenient matrix
$K(U_{\rm L}, U_{\rm R}, \phi )$~\cite{CCWZ}
is defined by the following transformation property of 
$\xi$ ($U = \xi^2$):
$\xi \rightarrow U_{\rm L} \cdot \xi \cdot K^{\dag} 
  = K \cdot \xi \cdot U_{\rm R}^{\dag}$,
and specifies the transformations of ``constituent-type'' objects.
The fields we need transform as
\begin{eqnarray}
&&
  N \rightarrow K \cdot N \cdot K^{\dag} \ ,
\nonumber\\
&&
  \rho_\mu \rightarrow K \cdot \rho_\mu \cdot K^{\dag}
  + \frac{i}{\tilde{g}} K \cdot \partial_\mu K^{\dag}
\ ,
\nonumber\\
&&
  F_{\mu\nu}(\rho) = 
  \partial_\mu \rho_\nu - \partial_\nu \rho_\mu - i 
  \tilde{g} \left[ \rho_\mu \,,\, \rho_\nu \right]
  \rightarrow
  K \cdot F_{\mu\nu} \cdot K^{\dag}
\ ,
\label{transf}
\end{eqnarray}
where the coupling constant $\tilde{g}$ is about $4.04$.
One may refer to Ref.~\cite{Harada-Schechter} for our treatment of the
pseudoscalar-vector Lagrangian and to
Ref.~\cite{Black-Fariborz-Sannino-Schechter:99} for the scalar
addition.
The entire Lagrangian is chiral invariant (modulo the quark mass term
induced symmetry breaking pieces) and, when
electromagnetism is added, gauge invariant.

It should be remarked that the effect of adding vectors to the chiral
Lagrangian of pseudoscalars only is to replace the photon coupling to
the charged pseudoscalars as,
\begin{eqnarray}
&&
i e {\cal A}_\mu \, 
\mbox{Tr}\left(
  Q \phi \mathop{\partial_\mu}^{\leftrightarrow} \phi 
\right)
\rightarrow
\nonumber\\
&& \quad
e {\cal A}_\mu \, 
\biggl[
  k \tilde{g} F_\pi^2 
  \mbox{Tr} \left( Q \rho_\mu \right)
\nonumber\\
&& \qquad
  + i \left( 1- \frac{k}{2} \right) 
  \mbox{Tr}\left(
    Q \phi \mathop{\partial_\mu}^{\leftrightarrow} \phi 
  \right)
\biggr]
+ \cdots
\ ,
\label{VMD}
\end{eqnarray}
where ${\cal A}_\mu$ is the photon field, 
$Q = \mbox{diag}(2/3,-1/3,-1/3)$ 
and $k = \left( \frac{m_v}{\tilde{g} F_\pi} \right)^2$
with $m_v \simeq 0.76\,\mbox{GeV}$.
The ellipses stand for symmetry breaking corrections.
We see that in this model, Sakurai's vector meson 
dominance~\cite{Sakurai} simply amounts to the 
statement that $k=2$ (the KSRF 
relation~\cite{KSRF}).
This is a reasonable numerical approximation which is essentially
stable to the addition of symmetry 
 breakers~\cite{Harada-Schechter,foot:VD} 
and we employ it here by neglecting the last term
in Eq.~(\ref{VMD}).
 Although vector meson dominance must be somewhat modified in
cases where the axial anomaly plays a role \cite{BKY:88}, it generally
works quite well for processes such as those we study here.

\begin{figure}[htbp]
\begin{center}
\epsfxsize = 6.5cm
\ \epsfbox{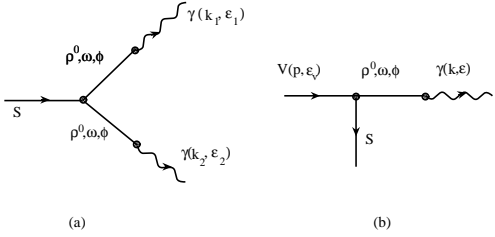}
\end{center}
\caption[]{%
Feynman diagrams for (a)~$S\rightarrow \gamma\gamma$ and
(b)~$V \rightarrow S \gamma$.
}\label{fig:1}
\end{figure}

The new feature of the present work is the inclusion of strong
trilinear scalar-vector-vector terms in the effective Lagrangian:
\begin{eqnarray}
&&
{\cal L}_{SVV} =  \beta_A \,
\epsilon_{abc} \epsilon^{a'b'c'}
\left[ F_{\mu\nu}(\rho) \right]_{a'}^a
\left[ F_{\mu\nu}(\rho) \right]_{b'}^b N_{c'}^c
\nonumber\\
&& \quad
{}+
 \beta_B \, \mbox{Tr} \left[ N \right]
\mbox{Tr} \left[ F_{\mu\nu}(\rho) F_{\mu\nu}(\rho) \right]
\nonumber\\
&& \quad
{}+
 \beta_C \, \mbox{Tr} \left[ N F_{\mu\nu}(\rho) \right]
\mbox{Tr} \left[ F_{\mu\nu}(\rho) \right]
\nonumber\\
&& \quad
{}+
 \beta_D \, \mbox{Tr} \left[ N \right]
\mbox{Tr} \left[ F_{\mu\nu}(\rho) \right]
\mbox{Tr} \left[ F_{\mu\nu}(\rho) \right]
\ .
\label{SVV}
\end{eqnarray}
Chiral invariance is evident from (\ref{transf}) and the four
flavor-invariants are needed for generality.  (A term 
$\sim \mbox{Tr}( FFN )$ is linearly dependent on the four shown).
Actually the $\beta_D$ term will not contribute in our model so there
are only three relevant parameters $\beta_A$, $\beta_B$ and $\beta_C$.
Equation~(\ref{SVV}) is analogous to the $PVV$ interaction which was
originally introduced as a $\pi\rho\omega$ coupling a long time 
ago~\cite{GSW}. It is intended to be a leading point-like
\cite{nna} description of the production mechanism.
With (\ref{VMD}) one can now compute the amplitudes for 
$S\rightarrow\gamma\gamma$ and $V \rightarrow S \gamma$ according to
the diagrams of Fig.~\ref{fig:1}.

The decay matrix element for $S \rightarrow \gamma \gamma$ is
written as 
$(e^2/\tilde{g}^2) X_S \times \left(
k_1\cdot k_2 \, \epsilon_1\cdot \epsilon_2 -
k_1\cdot \epsilon_2 \, k_2 \cdot \epsilon_1
\right)$ 
where $\epsilon_\mu$ stands for the photon polarization vector.  It is
related to the width by
\begin{equation}
\Gamma \left( S \rightarrow \gamma\gamma \right)
=
\alpha^2 \frac{\pi}{4} m_S^3
\left\vert \frac{X_S}{\tilde{g}^2} \right\vert^2
\ ,
\label{swidth}
\end{equation}
and $X_S$ takes on the specific forms:
\begin{eqnarray}
X_{\sigma} &=&
  \frac{4}{9} \beta_A 
    \left( \sqrt{2} s - 4 c \right)
  {} + \frac{8}{3} \beta_B
    \left( c - \sqrt{2} s \right)
\ ,
\nonumber\\
X_{f_0} &=&
  - \frac{4}{9} \beta_A 
    \left( \sqrt{2} c + 4 s \right)
  {} + \frac{8}{3} \beta_B
    \left( \sqrt{2} c + s \right)
\ , 
\nonumber\\
X_{a_0} &=& \frac{4\sqrt{2}}{3} \beta_A 
\ .
\label{samps}
\end{eqnarray}
Here $\alpha=e^2/(4\pi)$, $s = \sin \theta_S$ and $c = \cos\theta_S$ where 
the scalar
mixing angle, $\theta_S$ is defined from
\begin{equation}
\left(\begin{array}{c}
\sigma \\ f_0
\end{array}\right)
=
\left(\begin{array}{cc}
c & -s \\ s & c
\end{array}\right)
\left(\begin{array}{c}
N_3^3 \\ (N_1^1 + N_2^2)/\sqrt{2}
\end{array}\right)
\ .
\label{mixing}
\end{equation}
Furthermore ideal mixing for the vectors, with
$\rho^0 = (\rho_1^1 - \rho_2^2)/\sqrt{2}$,
$\omega = (\rho_1^1 + \rho_2^2)/\sqrt{2}$,
$\phi = \rho_3^3$, was assumed for simplicity.

Similarly, the decay matrix element for $V\rightarrow S \gamma$ is
written as $(e/\tilde{g}) C_V^S \times \left[
p\cdot k \epsilon_V \cdot \epsilon 
- p \cdot \epsilon k \cdot \epsilon_V
\right]$.
It is related to the width by
\begin{equation}
\Gamma (V \rightarrow S \gamma ) =
\frac{\alpha}{3} \left\vert k_V^S \right\vert^3
\left\vert \frac{C_V^S}{\tilde{g}} \right\vert^2
\ ,
\end{equation}
where $k_V^S = (m_V^2 - m_S^2)/(2m_V)$ is the photon momentum in the
$V$ rest frame.
For the energetically allowed $V \rightarrow S \gamma$ processes we
have
\begin{eqnarray}
C_\phi^{f_0} &=& 
 \frac{2\sqrt{2}}{3} \beta_A c
 - \frac{4}{3} \beta_B \left( \sqrt{2} c + s \right)
\nonumber\\
&& \quad
 + \frac{\sqrt{2}}{3} \beta_C
   \left( c - \sqrt{2} s \right)
\ , 
\nonumber\\
C_\phi^{\sigma} &=& 
 - \frac{2\sqrt{2}}{3} \beta_A s
 - \frac{4}{3} \beta_B 
   \left( c - \sqrt{2} s \right)
\nonumber\\
&& \quad
 - \frac{2}{3} \beta_C
   \left( c + \frac{1}{\sqrt{2}} s \right)
\ , 
\nonumber\\
C_\phi^{a_0} &=& \sqrt{2} \left( \beta_C - 2 \beta_A \right)\ , 
\nonumber\\
C_\omega^{\sigma} &=& 
 \frac{2\sqrt{2}}{3} \beta_A 
   \left( c + \sqrt{2} s \right)
 + \frac{2\sqrt{2}}{3} \beta_B 
   \left( c - \sqrt{2} s \right)
\nonumber\\
&& \quad
 - \frac{2}{3} \beta_C
   \left( \sqrt{2} c + s \right),
\nonumber\\
C_{\rho^0}^\sigma &=& -2\sqrt{2} \beta_A c + 2 \sqrt{2} \beta_B
\left( c- \sqrt{2} s \right).
\label{vamps}
\end{eqnarray}

In addition, the same model predicts amplitudes for
 the energetically allowed
$S\rightarrow V \gamma$ processes:
$f_0 \rightarrow \omega \gamma$, $f_0 \rightarrow \rho^0 \gamma$,
$a_0^0 \rightarrow \omega \gamma$,
$a_0^0 \rightarrow \rho^0 \gamma$ and,
if $\kappa^0$ is sufficiently heavy 
$\kappa^0 \rightarrow K^{\ast0} \gamma$.
The corresponding width is
\begin{equation}
\Gamma (S \rightarrow V \gamma ) =
 \alpha \left\vert k_S^V \right\vert^3
\left\vert \frac{D_S^V}{\tilde{g}} \right\vert^2
\ ,
\end{equation}
where $k_S^V = (m_S^2 - m_V^2)/(2m_S)$ and
\begin{eqnarray}
D_{f_0}^\omega &=& 
 \frac{2}{3} \beta_A \left( - 2 c +\sqrt{2} s \right)
 + \frac{2}{3} \beta_B \left( 2 c +\sqrt{2} s \right)
\nonumber\\
&& \quad
 + \frac{2}{3} \beta_C
   \left( c - \sqrt{2} s \right)
\ , 
\nonumber\\
D_{f_0}^{\rho^0} &=& 
 - 2 \sqrt{2} \beta_A s
 + 2 \beta_B 
   \left(2 c +\sqrt{2} s \right)
\ , 
\nonumber\\
D_{a_0}^\omega &=& 2 \beta_C \ ,
\nonumber\\
D_{a_0}^{\rho^0} &=&  \frac{4}{3} \beta_A \ ,
\nonumber\\
D_{\kappa^0}^{K^{\ast0}} &=& 
 - \frac{8}{3} \beta_A 
\ .
\label{svgamps}
\end{eqnarray}

All the different decay amplitudes are described by the parameters
$\beta_A$, $\beta_B$ and $\beta_C$.
The reason $\beta_D$ does not appear at all and $\beta_C$ does not
appear for $S \rightarrow \gamma \gamma$ is that, noting
Eq.~(\ref{VMD}), the $\mbox{Tr} ( F_{\mu\nu} )$ factor is seen to give
zero when coupled to an external photon line.
Because the $\sigma$ and $\kappa$ are so broad, the simple two body
final state approximation in decays like 
$\omega$, $\phi \rightarrow \sigma \gamma \rightarrow \pi^0 \pi^0
\gamma$ is not accurate.  It is better to consider these decays as
having three body final states with the terms in Eq.~(\ref{SVV}) giving
the vertices and to take into account large width corrections in the
scalar propagators as well as non resonant background.

These formulas can be used for different choices of the quark
structure of the scalar nonet $N_a^b$ (e.g. the usual $q_a \bar{q}^b$
scenario or the ``dual'' scenario $Q_a \bar{Q}^b$ where 
$Q_a \sim \epsilon_{abc} \bar{q}^b \bar{q}^c$).
The characteristic mixing angle $\theta_S$ is expected to differ,
depending on the scheme.  In the literature, besides conventional 
$q \bar{q}$ models, $qq \bar{q}\bar{q}$ 
models~\cite{Jaffe},
meson-meson ``molecule'' models~\cite{IW}
and unitarized meson-meson~\cite{umm} 
models have been investigated.
Recently models featuring mixing between a $qq \bar{q}\bar{q}$ nonet
and a heavier $q \bar{q}$ nonet have been proposed~\cite{mixing};
in this case two sets of interactions like Eq.~(\ref{SVV}) should be
included.

Now we shall illustrate the procedure for the model of a single
putative scalar nonet~\cite{Black-Fariborz-Sannino-Schechter:99} with
a mixing angle, $\theta_S \simeq - 20^{\circ}$ (characteristic of
$qq{\bar q}{\bar q}$ type scalars).

The parameters $\beta_A$ and $\beta_B$ may be estimated from the
$S\rightarrow \gamma\gamma$ processes.  Substituting 
$\Gamma_{\rm exp}(a_0\rightarrow \gamma\gamma) =
(0.28\pm0.09)\,\mbox{keV}$ (obtained using \cite{PDG}
$B(a_0 \rightarrow K \bar{K})/B( a_0 \rightarrow \eta \pi) =
0.177 \pm 0.024$)
into Eqs.~(\ref{swidth}) and (\ref{samps})
yields $\beta_A = (0.72\pm0.12)\,\mbox{GeV}$
(assumed positive in sign).
Of course, this value is independent of the value of $\theta_S$.
Then, $\Gamma_{\rm exp}(f_0 \rightarrow \gamma\gamma) = 
0.39\pm0.13\,\mbox{keV}$ yields either 
$\beta_B = (0.61\pm0.10)\,\mbox{GeV}^{-1}$ or
$\beta_B = (-0.62\pm0.10)\,\mbox{GeV}^{-1}$.
In turn we formally predict $\Gamma(\sigma\rightarrow \gamma\gamma)$
to be either $(0.024\pm0.023)\,\mbox{keV}$ or
$(0.38\pm0.09)\,\mbox{keV}$ respectively.

Next consider the $\phi$ radiative decays.  Assuming $\phi\rightarrow
\eta\pi^0\gamma$ is dominated by $\phi\rightarrow a_0\gamma$, 
$\Gamma_{\rm exp}(\phi \rightarrow a_0 \gamma) =
(0.47\pm0.07)\,\mbox{keV}$ and Eq.~(\ref{vamps}) determines $\beta_C$
as either $(7.7\pm0.5)\,\mbox{GeV}^{-1}$ or
$(-4.8\pm0.5)\,\mbox{GeV}^{-1}$.
Note that $\vert\beta_A\vert$ and $\vert\beta_B\vert$ are almost an
order of magnitude smaller than $\vert\beta_C\vert$.  Thus, the $\phi$
radiative decay rates are mainly determined by $\vert\beta_C\vert$.
Knowing $\beta_A$, $\beta_B$ and $\beta_C$ we can predict 
$\Gamma(\phi\rightarrow f_0\gamma)$ using Eq.~(\ref{vamps}).
There are four possibilities due to the two possibilities each for
$\beta_B$ and $\beta_C$.  The largest number, 
$\Gamma(\phi\rightarrow f_0\gamma) = (0.21\pm0.03)\,\mbox{keV}$ 
corresponds \cite{explainphasespace} to the choice $\beta_B = 
(-0.62\pm0.10)\,\mbox{GeV}^{-1}$
and $\beta_C = (7.7\pm0.5)\,\mbox{GeV}^{-1}$.

Unfortunately this is still considerably smaller than
the listed value~\cite{PDG}:
$\Gamma_{\rm exp}(\phi\rightarrow f_0\gamma) =
(1.51\pm 0.41)\,\mbox{keV}$%
~\cite{foot:phif0gam}.
Recently Close and Kirk \cite{Close-Kirk} proposed
that the ratio $\Gamma(\phi \rightarrow f_0\gamma)/\Gamma(\phi
\rightarrow a_0\gamma)$ 
  could be boosted by considering the effects of the
isospin violating $a_0^0(980)$-$f_0(980)$ mixing.  We will now
see that these effects are small in our model. 
  One may simply introduce the mixing
by a term in the effective Lagrangian:
${\cal L}_{af} = A_{af} a_0^0 f_0$.
A recent calculation~\cite{ABFS} for the purpose of finding the effect
of the scalar mesons in the $\eta\rightarrow 3\pi$ process obtained
the value
$A_{af} = -4.66\times10^{-3}\,\mbox{GeV}^2$.
It is convenient to treat this term as a perturbation.
  Then the
amplitude for $\phi\rightarrow f_0\gamma$ includes a correction term
consisting of the $\phi\rightarrow a_0^0\gamma$ amplitude given in 
Eq.~(\ref{vamps}) multiplied by $A_{af}$ and by the $a_0$ propagator.
The $\phi\rightarrow a_0^0\gamma$ amplitude has a similar correction.
In terms of the amplitudes in Eq.(\ref{vamps}) the desired ratio is then,
\begin{equation}
\frac{amp(\phi\rightarrow f_0\gamma)}{amp(\phi\rightarrow a_0^0\gamma)}=
\frac{ C_\phi^f + A_{af}C_\phi^a/D_a(m_f^2) }
{ C_\phi^a + A_{af}C_\phi^f/D_f(m_a^2)},
\label{amprat}
\end{equation}
where $D_a(m_f^2)=-m_f^2 + m_a^2 -im_a\Gamma_a$ and $D_f(m_a^2) =
-m_a^2 +m_f^2 -im_f\Gamma_f$. In this approach the propagators are diagonal in the 
isospin basis.
 The numerical values of these
resonance widths and masses are,
 according to the Review of
Particle Physics~\cite{PDG} $m_{a_0} = (984.7\pm1.3)\,\mbox{MeV}$,
$\Gamma_{a_0} = 50$--$100$\,MeV, $m_{f_0} = 980 \pm10\,\mbox{MeV}$
and $\Gamma_{f_0} = 40$--$100$\,MeV. For definiteness, from 
 column 1 of Table II in Ref.~\cite{Harada-Sannino-Schechter} we
take $m_{f_0} = 987\,\mbox{MeV}$ and $\Gamma_{f_0} = 65\,\mbox{MeV}$
while in Eq.~(4.2) of Ref.~\cite{FS}
we take $\Gamma_{a_0} = 70 \,\mbox{MeV}$.
In fact the main conclusion does not depend on these precise values. 
It is easy to see that the mixing factors are approximately 
given by
\begin{equation}
\frac{A_{af}}{D_a(m_f^2)} \approx \frac{A_{af}}{D_f(m_a^2)}
\approx \frac{iA_{af}}{m_a\Gamma_a} \approx -0.07i.
\label{mfapprox}
\end{equation} 
Noting that $C_\phi^f/C_\phi^a \approx 0.75 $ in the present model,
the ratio in Eq.(\ref{amprat}) is roughly $(0.75-0.07i)/(1-0.05i)$.
Clearly, the correction to $\Gamma(\phi \rightarrow f_0\gamma)/\Gamma(
\phi \rightarrow a_0\gamma)$ due to $a_0$-$f_0^0$ mixing only amounts
to a  few per cent, nowhere near the huge effect suggested 
in \cite{Close-Kirk}.
   It may be remarked that Eq.(\ref{amprat}) is practically accurate to 
all orders in $A_{af}$, corresponding to iterating any number
of $a_0$-$f_0$ transitions. Then, after summing a geometric series,
the numerator picks up a correction factor $[1-A_{af}^2/
(D_a(m_f^2)D_f(m_f^2))]^{-1}$ and the denominator, the similar
factor $[1-A_{af}^2/(D_a(m_a^2)D_f(m_a^2))]^{-1}$.

    Vector meson dominance, together with the assumptions of $SU(3)$
flavor symmetry and a single nonet of scalar mesons makes many more 
predictions. These are listed in Table I for two of the allowed $
(\beta_A, \beta_B, \beta_C)$ parameter sets, neglecting $a_0$-$f_0$ mixing.
It will be interesting to see if future experiments confirm the pattern 
of predicted widths.

\begin{table}[htbp]
\begin{center}
\begin{tabular}{lrr}
\hline
$\beta_A$ & $0.72\pm0.12$ & $0.72\pm0.12$ \\
$\beta_B$ & $0.61\pm0.10$ & $-0.62\pm0.10$ \\
$\beta_C$ & $7.7\pm0.52$ & $7.7\pm0.52$ \\
\hline
$f_0/a_0$ ratio & $0.26\pm0.06$ & $0.46\pm0.09$ \\
$\Gamma(\sigma\rightarrow\gamma\gamma)$ 
   & $0.024\pm0.023$ & $0.38\pm0.09$ \\
$\Gamma(\phi\rightarrow\sigma\gamma)$ & $137\pm19$ & $33\pm9$ \\
$\Gamma(\omega\rightarrow\sigma\gamma)$ & $16\pm3$ & $33\pm4$ \\
$\Gamma(\rho\rightarrow\sigma\gamma)$ & $0.23\pm0.47$ & $17\pm4$ \\
$\Gamma(f_0\rightarrow\omega\gamma)$ 
  & $126\pm20$ & $88\pm17$ \\
$\Gamma(f_0\rightarrow\rho\gamma)$ & $19\pm5$ & $3.3\pm2.0$ \\
$\Gamma(a_0\rightarrow\omega\gamma)$ & $641\pm87$ & $641\pm87$ \\
$\Gamma(a_0\rightarrow\rho\gamma)$ & $3.0\pm1.0$ & $3.0\pm1.0$ \\
\hline
\end{tabular}
\end{center}
\caption[]{Fitted values of $\beta_A$, $\beta_B$ and $\beta_C$
together with the predicted values of the ratio 
$\Gamma(\phi\rightarrow f_0\gamma)/\Gamma(\phi\rightarrow a_0\gamma)$
and the decay widths of 
$V \rightarrow S + \gamma$ and $S \rightarrow V + \gamma$.
Units of $\beta_A$, $\beta_B$ and $\beta_C$ are $\mbox{GeV}^{-1}$
and those of the decay widths are keV.
}\label{tab1}
\end{table}

We have given a leading order correlation of many radiative decays 
involving 
scalars, based on flavor symmetry and vector meson dominance. Clearly
further improvements can be made.
    Elsewhere, we will study flavor symmetry breaking effects,
higher drivative interaction terms, treatment 
of the $S\gamma$ final states as $PP\gamma$, 
and the case of mixed $q{\bar q}$ and $qq{\bar q}{\bar q}$
scalar nonets.

We are happy to thank N. N. Achasov for a suggested correction
to the first version of this note and with
A. Abdel-Rahiem and A. H. Fariborz for very
helpful discussions.
D.B. wishes to acknowledge support from the Thomas Jefferson National
Accelerator Facility operated by the Southeastern Universities Research
Association (SURA) under DOE contract number DE-AC05-84ER40150.
The work of M.H. 
is supported in part by Grant-in-Aid for Scientific Research
(A)\#12740144 and USDOE Grant Number DE-FG02-88ER40388. The
work of J.S. is supported in part by DOE contract DE-FG-02-85ER40231.


\begin{thebibliography}{99}
\bibitem{explainphasespace}
We took $m_{f_0} = 987$ MeV here;
taking 980 MeV would almost double the result.
\bibitem{foot:phif0gam}
This value is obtained by assuming that 
$\phi \rightarrow \pi^0 \pi^0\gamma$ is dominated 
by $\phi \rightarrow f_0 \gamma \rightarrow \pi^0 \pi^0 \gamma$.
In Ref.~\cite{Achasov-Ivanchenko}
it was noted that a very small
contribution to decays like $\phi \rightarrow \pi^0 \pi^0\gamma$
 is obtained using two $VVP$ vertices and vector
meson dominance via $\phi \rightarrow \pi^0 \rho^0 \rightarrow
\pi^0 \pi^0 \omega \rightarrow \pi^0 \pi^0 \gamma$.  This is usually
called the ``vector meson dominance contribution'' in the literature.
Of course it exists in our model too but we shall neglect it here; 
we assume that our effective $VVS$ vertex describes the leading 
contribution.
\end{thebibliography}
\end{document}